\documentclass[pre, reprint, superscriptaddress]{revtex4-2}

\usepackage{graphicx}
\usepackage{dcolumn}
\usepackage{bm}
\usepackage[dvipsnames]{xcolor}
\usepackage{amsmath}
\usepackage[T1]{fontenc}

\begin{document}


\title{On the emergence of molecular tilt in a ferroelectric smectic liquid crystal with broken director-inversion symmetry}

\author{Aitor Erkoreka}
    \email{Corresponding author: ai.erkoreka@ehu.eus}
\affiliation{Department of Physics, Faculty of Science and Technology, University of the Basque Country EHU, Bilbao, Spain}

\author{Mauricio Vera-Arévalo}
\author{Alberto Concellón}
\affiliation{Departamento de Química Orgánica, Facultad de Ciencias, Universidad de Zaragoza, Zaragoza, Spain}
\affiliation{Instituto de Nanociencia y Materiales de Aragón (INMA), CSIC-Universidad de Zaragoza, Zaragoza, Spain}

\author{Sergio Diez-Berart}
\affiliation{GRPFM, Departament de Física, ETSEIB, Universitat Politècnica de Catalunya, Barcelona, Spain}
\author{Jordi Sellarès}
\affiliation{DILAB, Departament de Física, ESEIAAT, Universitat Politècnica de Catalunya, Terrassa, Spain}
\author{Adrià Gràcia-Condal}
\affiliation{GCM, Departament de Física i Centre de Recerca en Ciència i Enginyeria Multiescala de Barcelona (CCEM), Universitat Politècnica de Catalunya, Barcelona, Spain}

\author{Ibon Alonso}
\author{Josu Martinez-Perdiguero}%
\affiliation{Department of Physics, Faculty of Science and Technology, University of the Basque Country EHU, Bilbao, Spain}

\date{\today}

\begin{abstract}
The origin of some mesophases of the ferroelectric nematic realm is not yet well understood. In this work we study the highly polar liquid crystal MIO, a close structural analogue of the prototypical ferroelectric nematogen DIO, which exhibits a ferroelectric smectic A to ferroelectric smectic C (SmA$_{\text{F}}$--SmC$_{\text{F}}$) phase transition. Calorimetric, dielectric and light-scattering experiments reveal that it is a second-order phase transition with mean-field behavior, and is driven by the softening of the tilt elastic constant accompanied by the divergence of the amplitude of the associated dielectric mode.
\end{abstract}

\maketitle

\newpage
\section{Introduction}
The spontaneous emergence of order from simple interacting units is one of the most remarkable phenomena in nature. From the formation of lipid membranes and protein assemblies in biological systems to crystal growth, colloidal superstructures, and nanostructured materials, self-assembly underpins the emergence of complexity across nature and condensed matter. Liquid crystals (LCs) represent a prime example of self-assembly whereby anisotropic molecules can exhibit different degrees of order within a fluid phase. The simplest example is the nematic (N) phase, in which the constituent molecules only exhibit orientational order along an arbitrary direction usually described by the unit vector $\mathbf{n}$. Since the microscopic interactions do not distinguish between either molecular end, a property known as head-to-tail invariance, the states $\mathbf{n}$ and $-\mathbf{n}$ are indistinguishable and the phase is apolar.

Recently, the ferroelectric nematic (N$_{\text{F}}$) phase was discovered, in which the dipoles of elongated highly polar molecules spontaneously align and give rise to a macroscopic polarization $\mathbf{P}$ along the director, thus breaking the aforementioned director-inversion symmetry \cite{mandle_nematic_2017, nishikawa_fluid_2017, sebastian_ferroelectric_2020, chen_first_2020, sebastian_ferroelectric_2022}. Polarization reversal measurements yield $|\mathbf{P}|$ values comparable to those of solid ferroelectrics ($\sim \mu$C/cm$^2$). This constitutes a groundbreaking discovery in the field of soft matter because, although some mesophases were known to exhibit ferroelectric behavior for decades, positional order, as in smectic phases, was thought to be a fundamental prerequisite for the emergence of long-range polar order \cite{takezoe_antiferroelectric_2010}. Moreover, even if ferroelectric order was present in these systems, director-inversion symmetry was always conserved. The discovery of the N$_{\text{F}}$ phase was soon followed by the identification of a great number of related polar mesophases lacking director-inversion symmetry, both nematic and smectic, making up the so-called ferroelectric nematic realm. These findings have opened new avenues of research in two main directions. On the one hand, the coupling of polar and orientational/positional order brings about a rich phenomenology of great scientific interest. This knowledge could then pave the way for unprecedented applications beyond the classical area of display technologies, among which the field of nonlinear and quantum optics stands out \cite{folcia_ferroelectric_2022, sultanov_tunable_2024, lovsin_ferroelectric_2026}. On the other hand, there is an active quest to unveil the phase-transition mechanisms occurring in these systems and achieve a molecular-level understanding of their behavior. In this context, while the origin of the N$_{\text{F}}$ phase remains a mystery, it is now well-established that the transition from the apolar N phase to an intermediate antiferroelectric splay-modulated nematic (N$_{\text{S}}$) phase is driven by the flexoelectric coupling between splay deformation and electric polarization \cite{mertelj_splay_2018, sebastian_ferroelectric_2020, medlerupnik_antiferroelectric_2025}. More recently, bend-flexoelectricity has been found to be responsible for a spontaneous chiral symmetry-breaking transition and the formation of a heliconical structure from achiral building blocks in the so-called ferroelectric twist-bend nematic (N$_{\text{TBF}}$) phase \cite{erkoreka_flexoelectricity_2026}.

In this paper, we tackle the question regarding the origin of the ferroelectric smectic C phase (SmC$_{\text{F}}$), a tilted smectic phase with spontaneous polarization along the director \cite{kikuchi_ferroelectric_2024, hobbs_polar_2024}. In particular, we demonstrate that the SmA$_{\text{F}}$--SmC$_{\text{F}}$ phase transition is second order in nature with mean-field behavior, and is driven by the softening of the tilt elastic constant and simultaneous divergence of the amplitude of the associated dielectric mode. The striking similarity between the chemical structure of the studied compound and other ferroelectric nematogens that do not exhibit these phases reveals the fine-tuning of the molecular interactions that promote or suppress them.

\section{Materials and methods}

\subsection{Material}

The investigated compound is almost identical to the prototypical ferroelectric nematogen DIO \cite{nishikawa_fluid_2017}, the difference being that, in the former case, the phenyl group next to the dioxane ring is monofluorinated instead of difluorinated. Therefore, we will hereinafter refer to it as MIO. It has a dipole moment of $\approx 9$ D mainly oriented along the molecular long axis \cite{kikuchi_fluid_2022}. The compound was synthesized by adapting literature methods \cite{kikuchi_fluid_2022} and recrystallized several times from a dichloromethane/methanol mixture to afford a highly enriched trans-isomer sample ($>99\%$ trans, see Fig. S1 in the Supplemental Material for the corresponding $^1$H NMR spectrum \cite{SM}). Its molecular structure and phase transition temperatures can be seen in Fig. \ref{fig:figure1}(a).

\subsection{Modulated differential scanning calorimetry (MDSC)}

Heat capacity measurements were performed using a DSC-Q100 calorimeter from TA-Instruments working in the modulated mode (MDSC). This technique, just like AC calorimetry, besides providing heat capacity data, simultaneously gives phase shift data $\phi$ that allow the determination of the order of the phase transitions. For first-order phase transitions, $\phi$ grows and can show a sharp peak, while for second-order phase transitions $\phi$ exhibits a dip or can even appear flat. Slow cooling- and heating-runs were performed in order to obtain precision calorimetric data. Measurements at a scanning rate of $0.04$ K/min presented in the paper were carried out with a modulation amplitude of $\pm 0.07$ K and a modulation period of $20$ s. A more detailed description of the MDSC technique can be found elsewhere \cite{sied_binary_2002, puertas_thermodynamic_2004}.

\subsection{Dielectric measurements}
The complex dielectric function $\varepsilon(f) = \varepsilon'(f)-i\varepsilon''(f)$ was measured in the frequency range 10 Hz-1 MHz. Measurements were performed with an Alpha-A impedance analyzer, setting the oscillator voltage to 0.03 V$_\mathrm{rms}$. The material was introduced into a 15 $\mu$m-thick sandwich glass cell with low-resistance ITO electrodes and parallel rubbing (EHC Co. Ltd, Japan) by capillary action in the N phase and the temperature during the measurements was controlled with a hot stage (Linkam). The sample was cooled from $130^{\circ}$C at $0.25$ K/min. The complex dielectric permittivity was determined by dividing the measured capacitance by the capacitance of the empty cell. The stray capacitance of the measurement circuit was carefully taken into account in all cases. For a quantitative analysis of the dielectric relaxation processes, the data were fitted to the Havriliak-Negami (HN) formula with a conductivity term:

\begin{equation}
    \varepsilon (f) = \sum_{k} \frac{\Delta \varepsilon_k}{\left[1+\left(i \frac{f}{f_k}\right)^{\alpha_k} \right]^{\beta_k}} + \varepsilon_{\infty} +\frac{\sigma}{\varepsilon_0(i\,2\pi f)^{\lambda}}\mathrm{,}\label{HN_eq}
\end{equation}

\noindent
where $\Delta\varepsilon_k$, $f_k$, $\alpha_k$ and $\beta_k$ are respectively the dielectric strength, relaxation frequency and broadness exponents of mode $k$, $\varepsilon_{\infty}$ is the high-frequency dielectric permittivity, $\sigma$ is a measure of the conductivity, and $\lambda$ is an exponent between 0 and 1.

Absolute values of the splay and bend elastic constants were obtained by means of the capacitance method in the N phase. A planar-to-homeotropic
Fréedericksz transition was induced by applying an ac signal with an Agilent Precision LRC meter E4980A. The capacitance of the sample was measured as a function of the applied voltage, which was varied from 0.01 V$_\mathrm{rms}$ to 5 V$_\mathrm{rms}$, with a delay time of 20 s between the application of the ac signal and the acquisition of the capacitance value. The splay and bend elastic constants were extracted from the fitting of the entire voltage dependence of the capacitance to the theory \cite{morris_measurements_1986}.

\subsection{Dynamic light scattering (DLS)}
Dynamic light scattering (DLS) experiments were performed using a frequency-doubled diode-pumped Nd:YAG laser (532 nm, 100 mW attenuated by $100\times$). Two single-photon APD detectors (Laser Components) and a digital correlator (LS Instruments) were employed to obtain the autocorrelation function of the scattered light intensity. A single-mode optical fiber with a GRIN lens was used to collect the scattered light within one coherence area. The direction and the polarization of the incoming and detected light were chosen to probe different fluctuation modes \cite{degennes_physics_1995}. The intensity autocorrelation function ($g_2$) was fitted to $g_2-1=2(1-j_{\text{D}})j_{\text{D}}g_1+j_{\text{D}}^2g_1^2+y_0$, where $j_{\text{D}}$ is the ratio between the intensity of light that is scattered inelastically and the total scattered intensity, and $g_1$ is a single exponential function $g_1=\exp{(-t/\tau)}$. The relaxation rate $1/\tau$ was attributed to the chosen eigenmode of orientational fluctuations with the wavevector $\mathbf{q}$ equal to the scattering vector $\mathbf{q}_{\text{s}}$. The scattered intensity of a given
mode was determined as a product $j_{\text{D}}I_{\text{tot}}$, where $I_{\text{tot}}$ was the total detected intensity. All experiments were done in 15 $\mu$m-thick planar glass cells (EHC Co. Ltd, Japan), adjusting in each case the cooling rate and measurement time.

\begin{figure}
\includegraphics[width=0.45\textwidth]{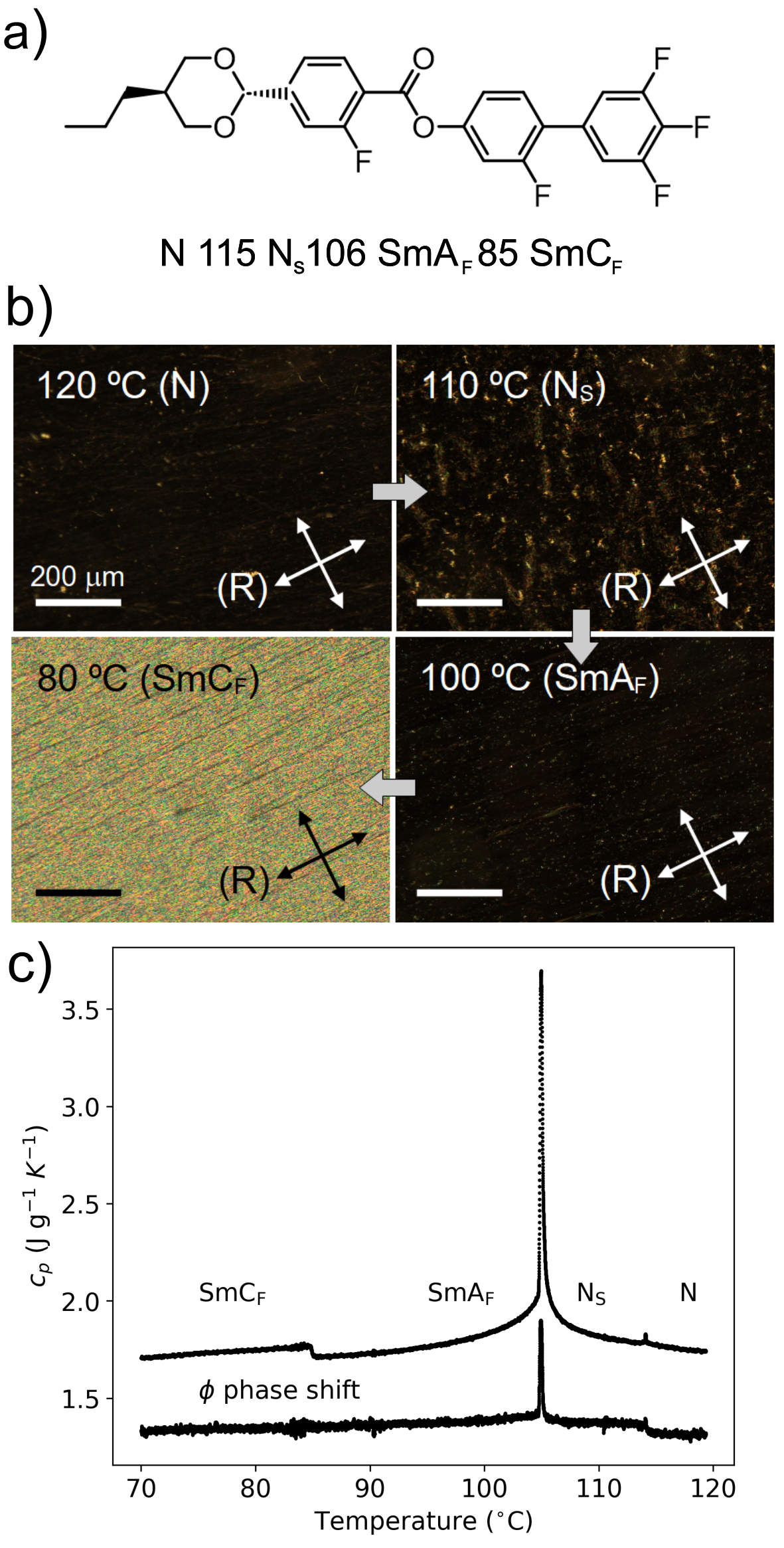}
\caption{\label{fig:figure1} (a) Molecular structure and phase transition temperatures of MIO. (b) Representative textures in a 15 $\mu$m-thick planar cell in the N, N$_{\text{S}}$, SmA$_{\text{F}}$ and SmC$_{\text{F}}$ phases. The scale, rubbing direction and polarizer/analyzer configuration are shown. (c) Heat-capacity curve obtained by MDSC at a heating rate of 0.04 K/min. Phase shift data are also shown.}
\end{figure}

\section{Results and discussion}

The minute difference in the fluorination pattern between the prototypical ferroelectric nematogen DIO and the studied material MIO alters the phase sequence from N--N$_{\text{S}}$--N$_{\text{F}}$ to N--N$_{\text{S}}$--SmA$_{\text{F}}$--SmC$_{\text{F}}$ \cite{nishikawa_fluid_2017, kikuchi_ferroelectric_2024}. This remarkable behavior seems to be intrinsic to the ferroelectric nematic realm, as in the well-studied case of RM734, another prototypical ferroelectric nematogen, in which the substitution of the terminal nitro group by a cyano group prevents the formation of the N$_{\text{F}}$ phase altogether \cite{mandle_molecular_2021}. This reveals the necessity of understanding the intricate molecular interactions occurring in these systems in order to achieve a profound understanding of mesophase behavior in the ferroelectric nematic realm. Fig. \ref{fig:figure1}(b) shows representative textures of MIO obtained by polarized optical microscopy (POM) in a 15 $\mu$m-thick planar cell across its entire phase sequence. While a high extinction ratio is observed in both the N and SmA$_{\text{F}}$ phases, suggesting a good alignment of the director $\mathbf{n}$, this is not true in the N$_{\text{S}}$ and SmC$_{\text{F}}$ phases, where extinction is completely lost in the latter case due to the development of the tilt. Our heat capacity measurements [Fig. \ref{fig:figure1}(c)] reveal that the transition from the N to the antiferroelectric splay-modulated N$_{\text{S}}$ phase is very weakly first order, as already quantified in DIO and RM734 \cite{thoen_phase_2023, thoen_calorimetric_2024}. The N$_{\text{S}}$--SmA$_{\text{F}}$ transition also seems to be weakly first-order but with a higher associated enthalpy change. Finally, the SmA$_{\text{F}}$--SmC$_{\text{F}}$ transition, of special interest to the present paper, is second-order, as deduced from the curve profile, the lack of hysteresis and the flat behavior of the phase shift. We will return our attention to this matter later.

\begin{figure}
\includegraphics[width=0.45\textwidth]{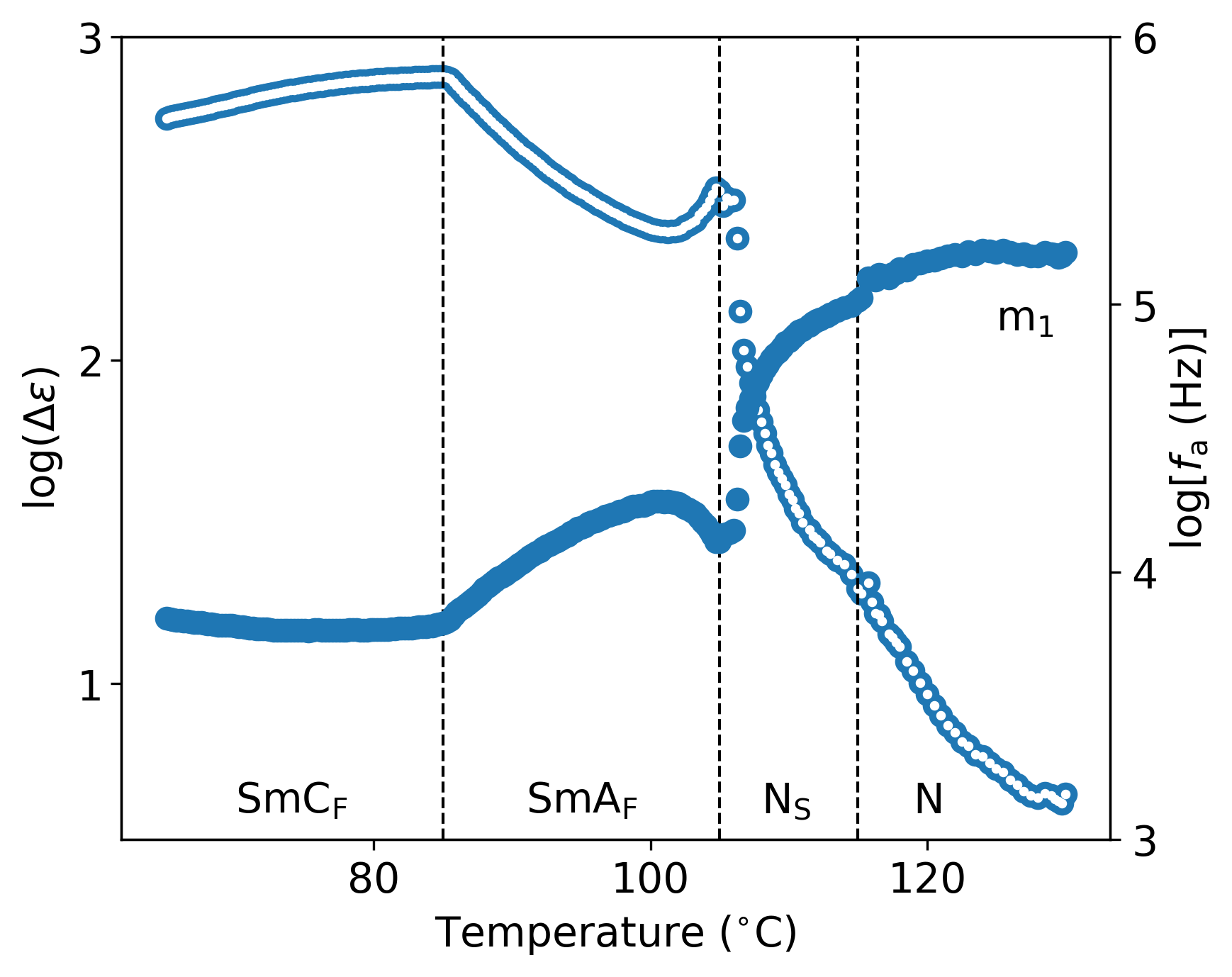}
\caption{\label{fig:figure2} Temperature evolution of the dielectric strengths ($\Delta \varepsilon$, empty symbols) and frequencies of maximum absorption ($f_{\text{a}}$, full symbols) of relaxation process m$_1$.}
\end{figure}

In order to probe the dipolar dynamics and polar correlations occurring in the different phases, we measured the dielectric spectra of MIO at different temperatures. These experiments were performed in a 15 $\mu$m-thick planar cell, so we measured the perpendicular component of permittivity $\varepsilon_{\perp}$. By fitting the spectra to Eq. (\ref{HN_eq}), we obtained the temperature evolution of the amplitude $\Delta \varepsilon$ and frequency of maximum absorption $f_{\text{a}}$ of the relevant dielectrically active relaxation processes (see Fig. S2 for fit examples \cite{SM}). The results are shown in Fig. \ref{fig:figure2}. The cell pseudo-relaxation due to the finite conductivity of the ITO layer was present at all temperatures but separated enough in frequency to allow an adequate deconvolution of the intrinsic relaxation processes. In the high-temperature N phase, m$_1$ is attributed to the rotation of molecules around their short axis, which becomes collective as the temperature is lowered. In RM734, this is the soft mode that leads to the N$_{\text{S}}$ and N$_{\text{F}}$ phases (which are very close in temperature) and exhibits a diverging amplitude and a critical slowing down characteristic of ferroelectric transitions \cite{sebastian_ferroelectric_2020, erkoreka_dielectric_2023}. In MIO, perhaps unsurprisingly due to their structural similarity, the situation is more similar to DIO, where there is no clear critical behavior \cite{erkoreka_collective_2023}. Nonetheless, m$_1$ does exhibit the typical soft-mode behavior at the N$_{\text{S}}$--SmA$_{\text{F}}$ transition, just like in DIO at the N$_{\text{S}}$--N$_{\text{F}}$ transition. However, it is important to point out that the dielectric behavior in the SmA$_{\text{F}}$ phase is markedly different from that of the N$_{\text{F}}$ phase. In fact, the dielectric behavior and permittivity values in the latter case have been a subject of intense debate \cite{clark_dielectric_2024, matko_interpretation_2024}. According to the interpretation proposed by some authors of the present paper, the colossal dielectric constants of the N$_{\text{F}}$ phase are caused by the behavior of the Goldstone mode (reorientation of spontaneous polarization) under confinement, which make the dielectric spectra become thickness-dependent \cite{erkoreka_dielectric_2023, erkoreka_collective_2023, erkoreka_constraining_2024}. This masks the true bulk response, and the low-temperature side of the transition to the N$_{\text{F}}$ phase has the characteristics described by the effective polarization-external capacitance Goldstone reorientation (PCG) model developed by Clark \textit{et al} \cite{clark_dielectric_2024}. This is not the case in the SmA$_{\text{F}}$ phase, where the rotational viscosity $\gamma_1$ is expected to be much larger and, thus, the PCG mechanism does not apply. Further down in temperature, m$_1$ again exhibits soft-mode behavior at the SmA$_{\text{F}}$--SmC$_{\text{F}}$ transition. It should be noted that in the N$_{\text{S}}$ phase, a lower-frequency process m$_2$ was identified. Its strong temperature dependence (see Fig. S3 \cite{SM}) suggests that m$_2$ is associated to some kind of collective dipolar fluctuation, but its origin is out of the scope of this paper.

\begin{figure}
\includegraphics[width=0.45\textwidth]{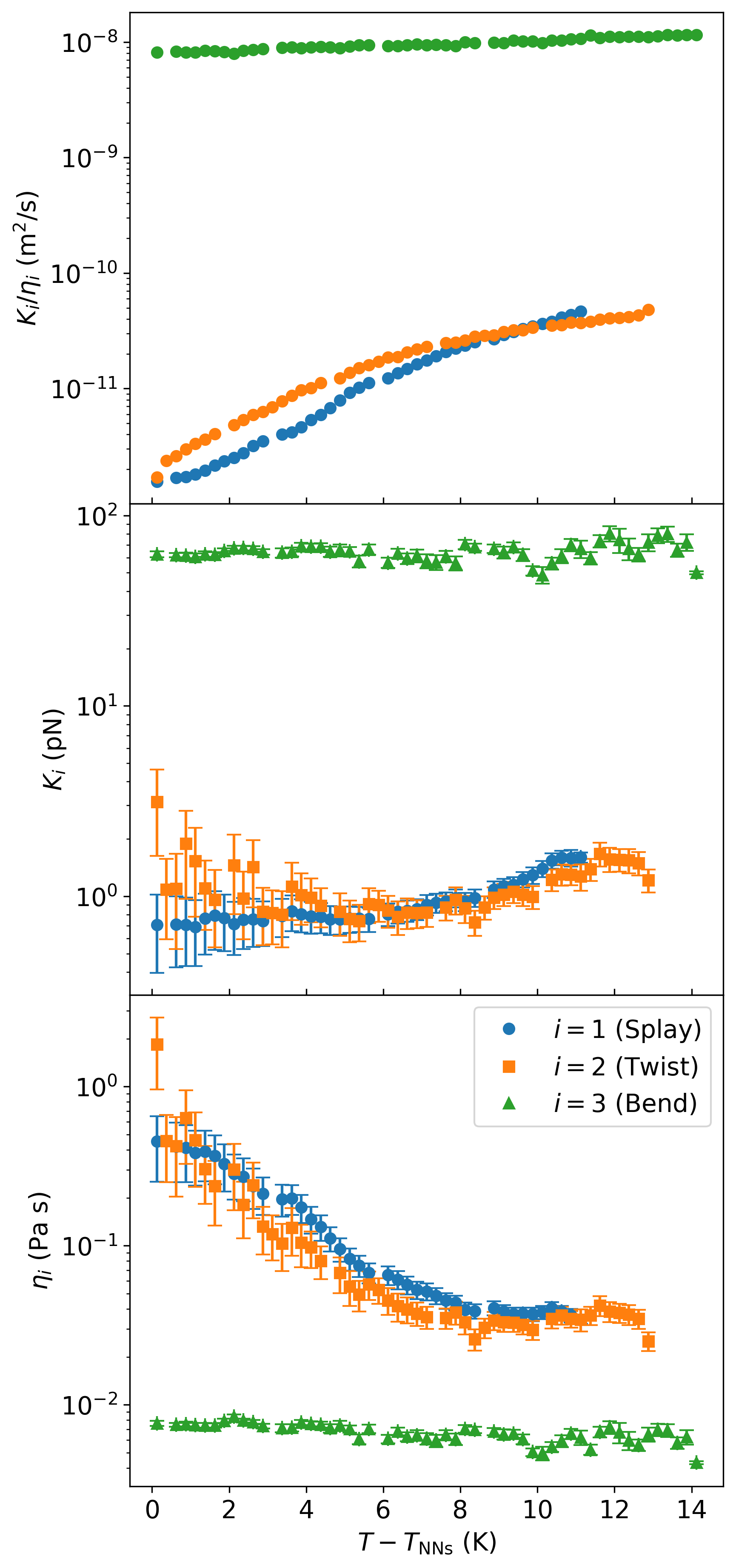}
\caption{\label{fig:figure3} Temperature evolution of the diffusivities ($K_i/\eta_i$, top), elastic constants ($K_i$, middle) and viscosities ($\eta_i$, bottom) in the N phase.}
\end{figure}

Before delving into the SmA$_{\text{F}}$--SmC$_{\text{F}}$ phase transition, we will analyze the transition from the N to the antiferroelectric splay-modulated N$_{\text{S}}$ phase. It is now a well-known fact that this transition is driven by the flexoelectric coupling between splay deformation and electric polarization \cite{mertelj_splay_2018, sebastian_ferroelectric_2020, medlerupnik_antiferroelectric_2025}, but the published data are scarce and the values and temperature evolution of the elastic constants in the N phase can reveal interesting insights. For this reason, we studied the evolution of orientational fluctuations by dynamic light scattering (DLS). These orientational fluctuations are the fundamental hydrodynamic excitations of the director field and have two dispersion branches: splay-bend and twist-bend \cite{degennes_physics_1995}. Their relaxation rates are proportional to a ratio of the nematic elastic constants $K$ and viscosity coefficients $\eta$. However, by selecting the appropriate scattering geometry, pure modes can be measured with relaxation rates $1/\tau=K_i q^2/\eta_i$ and intensities $I \propto (\Delta \varepsilon_{\text{opt}})^2/K_i q^2$, where $i=1,2,3$ denote splay, twist, and bend, and $q$ is the scattering vector \cite{degennes_physics_1995}. While twist viscosity equals rotational viscosity $\eta_2=\gamma_1$, the values
of bend and splay viscosities are affected by backflow, $\eta_1=\gamma_1-\alpha_3^2/\eta_\text{b}$ and $\eta_3=\gamma_1-\alpha_2^2/\eta_{\text{c}}$, where $\alpha_i$ are the Leslie viscosity coefficients and $\eta_\text{b,c}$ are Miesowicz viscosities \cite{degennes_physics_1995}. In order to obtain the temperature dependence of the anisotropy of the dielectric tensor at optical frequencies $\Delta \varepsilon_{\text{opt}}$, we measured the birefringence $\Delta n$ with a Berek compensator. Following this method we can only obtain the temperature dependence of the elastic constants but not their absolute values. For this purpose, we determined the absolute values of $K_1$ and $K_3$ at a reference temperature from the measurement of the Fréedericksz transition (see Figs. S4 and S5 \cite{SM}). $K_2$ was obtained from the ratio of $K_1/\eta_1$ and $K_2/\eta_2$ assuming that, in the case of splay, the backflow is negligible. The obtained diffusivities ($K_i/\eta_i$), elastic constants ($K_i$) and viscosities ($\eta_i$) can be found in Fig. \ref{fig:figure3}. We can see that, while the bend diffusivity stays practically constant as the temperature is lowered, both splay and twist diffusivites strongly decrease close to the transition to the N$_{\text{S}}$ phase. However, while in the twist case this is mainly due to the pretransitional increase in $\gamma_1$, for splay it also comes from the softening of $K_1$. In particular, $K_1$ reduces its value from 1.6 pN at $126^{\circ}$C to around 0.7 pN right before the transition. The shape of this temperature dependence is, as expected, more similar to DIO \cite{zhou_stereoisomer_2022, gleeson_freedericksz_2025, ghimire_dynamics_2025} than to RM734 \cite{mertelj_splay_2018}, where the softening happens closer to the phase transition temperature. In any case, we do not observe a pretransitional increase in $K_1$ which was identified in DIO \cite{zhou_stereoisomer_2022, gleeson_freedericksz_2025, ghimire_dynamics_2025}. $K_2$, on the other hand, is similar in value to $K_1$ but stays more or less constant and increases close to the N$_{\text{S}}$ phase. Finally, $K_3$ also stays more or less constant, but has a remarkably high value of $\sim 60$ pN. In RM734, its value was $\sim 12$ pN \cite{mertelj_splay_2018} while in DIO it was $\sim 6$ pN \cite{zhou_stereoisomer_2022, gleeson_freedericksz_2025, ghimire_dynamics_2025}. This is a very interesting observation and we suggest it is related to the stiffness of the MIO molecule, as well as to the presence of strong intermolecular correlations in the system, which strongly disfavor bend distortions. It should then come as no surprise that the system tends towards the formation of smectic phases. We will discuss how this can also be related with the development of tilt in the SmC$_{\text{F}}$ phase later.

\begin{figure}
\includegraphics[width=0.45\textwidth]{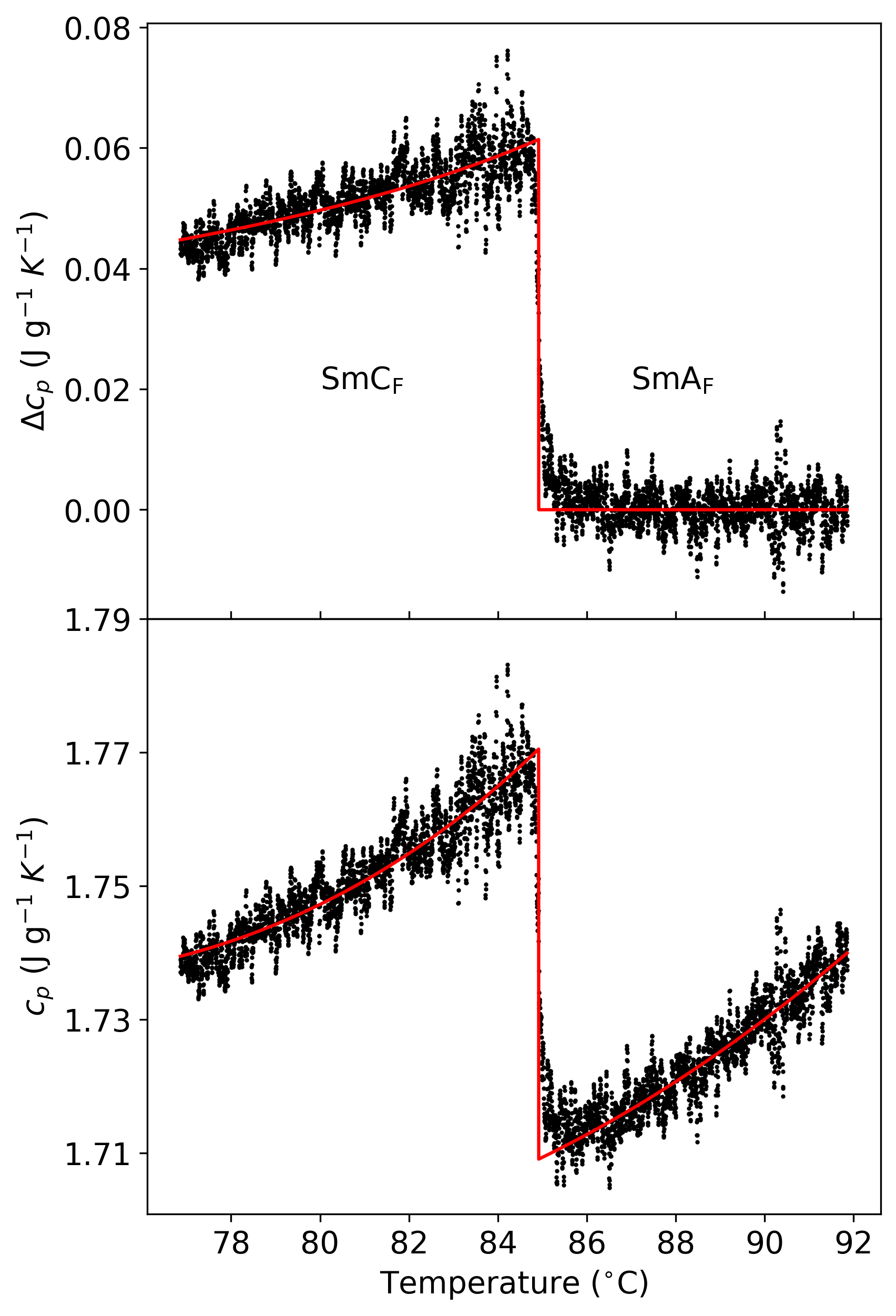}
\caption{\label{fig:figure4} Critical part of the specific heat (top) and total specific heat (bottom) at the SmA$_{\text{F}}$--SmC$_{\text{F}}$ phase transition. The red line at the top corresponds to the fit to Eq. (\ref{deltacp}), while at the bottom the background has been added.}
\end{figure}

In what follows, we will focus on the SmA$_{\text{F}}$--SmC$_{\text{F}}$ transition. Regarding the classical SmA--SmC transition, Pierre G. de Gennes predicted that it should be continuous (second-order) and exhibit helium-like critical behavior \cite{degennes_physics_1995}. He introduced the bidimensional order parameter $\psi=\omega e^{i\varphi}$, where $\omega$ is the tilt angle of the molecules and $\varphi$ defines their azimuthal position. Of course, $\langle |\psi| \rangle$ is only nonzero in the SmC phase, and $\langle |\psi|^2 \rangle$ is nonzero in the SmA phase and diverges at the transition. In any case, many works reported mean-field behavior and there were some theoretical arguments to explain this on the basis of a reduction of the critical regime \cite{degennes_physics_1995}. In the present case, we can draw a direct analogy since the SmA$_{\text{F}}$ and SmC$_{\text{F}}$ phases are ferroelectric equivalents of the SmA and SmC phases. If we assume that the magnitude of spontaneous polarization is already saturated, then the transition can be described in the same way, remembering that the development of the tilt leads to a secondary order parameter $\mathbf{P}_{\perp}$, the component of spontaneous polarization perpendicular to the layer normal (i.e. in the plane of the layers). Of course, $|\mathbf{P}_{\perp}|\propto |\psi|$ and they are interchangable. In addition, mean-field behavior would be expected in the present case since long-range dipolar interactions in ferroelectrics typically suppress critical fluctuations \cite{nattermann_ferroelectric_1975, yeomans_statistical_1992}. Indeed, this second-order mean-field nature can be identified in our precision calorimetry measurements at the transition shown in Fig. \ref{fig:figure4}. Substraction of the background specific heat through a second-order polynomial allows the extraction of the critical component. This behavior can be nicely explained with a simple Landau expansion of the free energy up to sixth order in $|\psi|$:

\begin{equation}
    F=at|\psi|^2+b|\psi|^4+c|\psi|^6,
\end{equation}
where $t=(T-T_{\text{AC}})/T_{\text{AC}}$, and $a$, $b$ and $c$ are positive constants. Minimizing the free energy with respect to the order parameter, it is possible to obtain the temperature dependence of the critical part of the heat capacity. The result is the following:

\begin{equation}
    \Delta c_p =  
\begin{cases}
A T (T_{\text{m}}-T)^{-1/2}, & T<T_{\text{AC}} \\
0, & T>T_{\text{AC}}
\end{cases}\label{deltacp}
\end{equation}
where $T_{\text{m}}=T_{\text{AC}}(1+t_0/3)$, $t_0=b^2/ac$ and $A=a^{3/2}/[2(3c)^{1/2}T_{\text{AC}}^{3/2}]$ \cite{huang_nature_1982, lien_possible_1984}. From the fit of our experimental data of Fig. \ref{fig:figure4} to Eq. (\ref{deltacp}) we obtain $A=(5.45\pm 0.04)\times 10^{-4}$ J g$^{-1}$ K$^{-3/2}$ and $T_{\text{m}}=368.2\pm0.2$ K. Taking $T_{\text{AC}}$ at the midpoint of the heat-capacity jump, then $T_{\text{AC}}=358.07$ K and, thus, $t_0=8.5\times 10^{-2}$. For compounds exhibiting the classical SmA--SmC transition, $t_0\sim 10^{-3}$, which is smaller in comparison with other systems showing mean-field transitions \cite{huang_nature_1982, lien_possible_1984}, including solid ferroelectrics like triglycine sulfate (TGS) \cite{strukov_heat_1965}. In fact, our value of $t_0$ is closer to that of the latter. Although the microscopic origin of $t_0$ is not clear, we speculate that the similarity between MIO and TGS may come from the presence of strong long-range dipolar interactions in these highly polar systems.

If the SmA$_{\text{F}}$--SmC$_{\text{F}}$ transition is driven by the softening of the tilt elastic constant, then $a t$ is evidently connected to the tilt elastic constant $D$, which goes to zero at the transition as $D\propto (T-T_{\text{AC}})$ in the mean-field case. As mentioned earlier, in the SmA$_{\text{F}}$ phase there will be strong fluctuations of the tilt angle close to the SmC$_{\text{F}}$ phase. These can of course be detected by DLS in the twist-bend branch. Recalling the result from the classical SmA--SmC transition, the intensity of this mode becomes \cite{delaye_critical_1976, brown_advances_1979, huang_light_1988}

\begin{equation}
    I_2 \propto \frac{(\Delta \varepsilon_{\text{opt}})^2}{D+K_2q_{\perp}^2+K_3q_{\parallel}^2}.
\end{equation}

\begin{figure}
\includegraphics[width=0.4\textwidth]{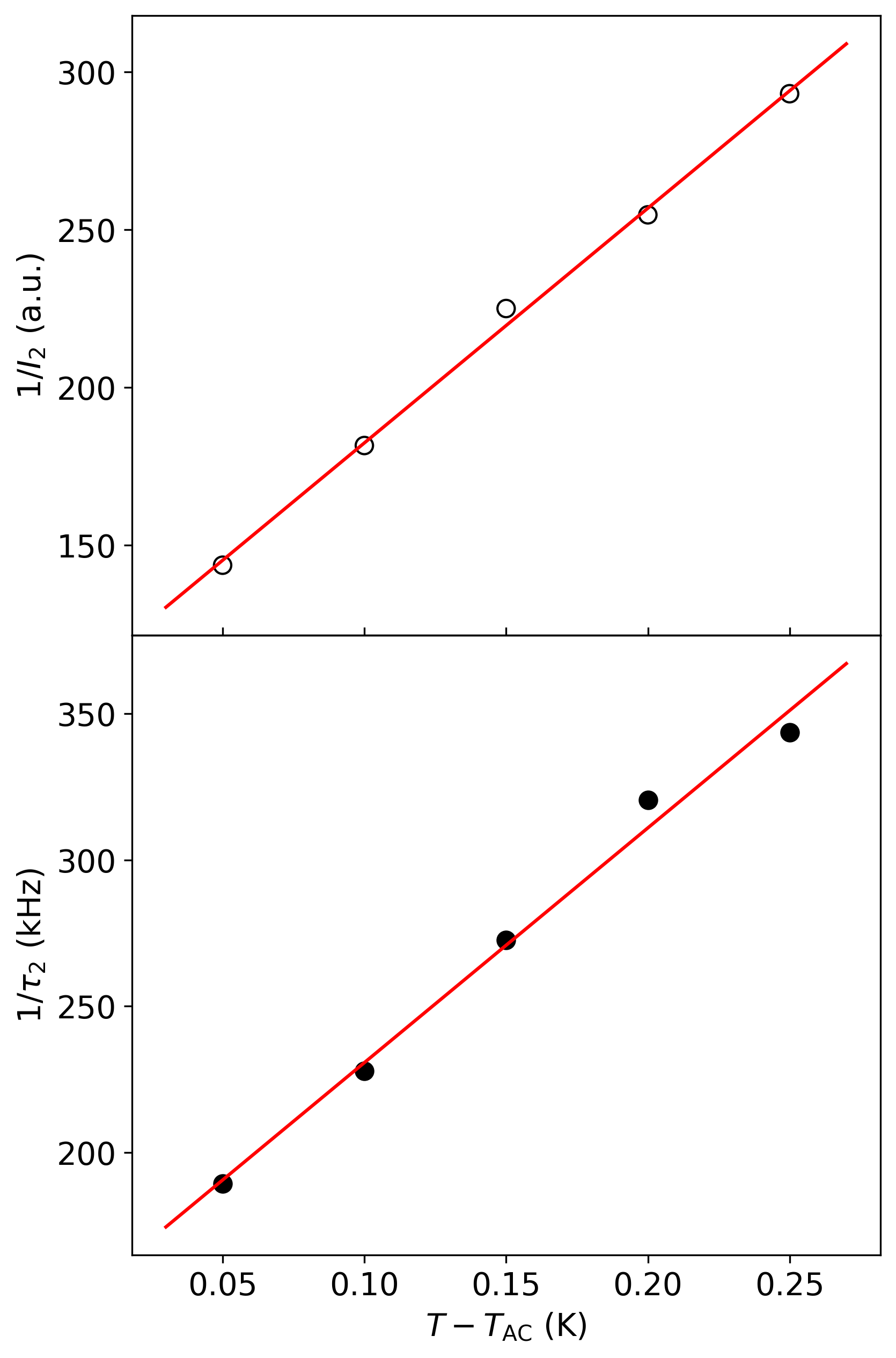}
\caption{\label{fig:figure5} Temperature dependence of the inverse of the scattering intensity ($1/I_2$, top) and relaxation rate ($1/\tau_2$, bottom) of the twist-bend mode above the SmA$_{\text{F}}$--SmC$_{\text{F}}$ transition. Red lines are linear fits to the experimental data.}
\end{figure}

If the scattering wavevector is small enough such that $q\, \xi <1$ and $K_2 q_{\perp}^2+K_3 q_{\parallel}^2 < D$, where $\xi$ is the coherence length of the fluctuations, then the scattering intensity reflects only the behavior of $D$ \cite{brown_advances_1979}. From our birefringence measurements (see Fig. S6 \cite{SM}), $\Delta \varepsilon_{\text{opt}}$ remains practically constant in the vicinity of the SmA$_{\text{F}}$--SmC$_{\text{F}}$ transition, so we should observe the behavior $1/I_2 \propto (T-T_{\text{AC}})$. In addition, the corresponding relaxation rate of this soft mode is $1/\tau=D/\eta$, where $\eta$ is a viscosity coefficient, so it follows analogous behavior $1/\tau \propto (T-T_{\text{AC}})$ if $\eta$ is regular. This is precisely what we observed in our measurements, shown in Fig. \ref{fig:figure5}. In the SmA$_{\text{F}}$ phase, far from the transition to the SmC$_{\text{F}}$ phase, $D$ is large in order to maintain the molecules normal to the smectic layers. Thus, the mode is fast and the scattered intensity low. Nonetheless, as the SmA$_{\text{F}}$--SmC$_{\text{F}}$ is approached, $D$ goes to zero and becomes observable in our DLS experiment very close to the phase transition temperature $T_{\text{AC}}$.

\begin{figure}
\includegraphics[width=0.4\textwidth]{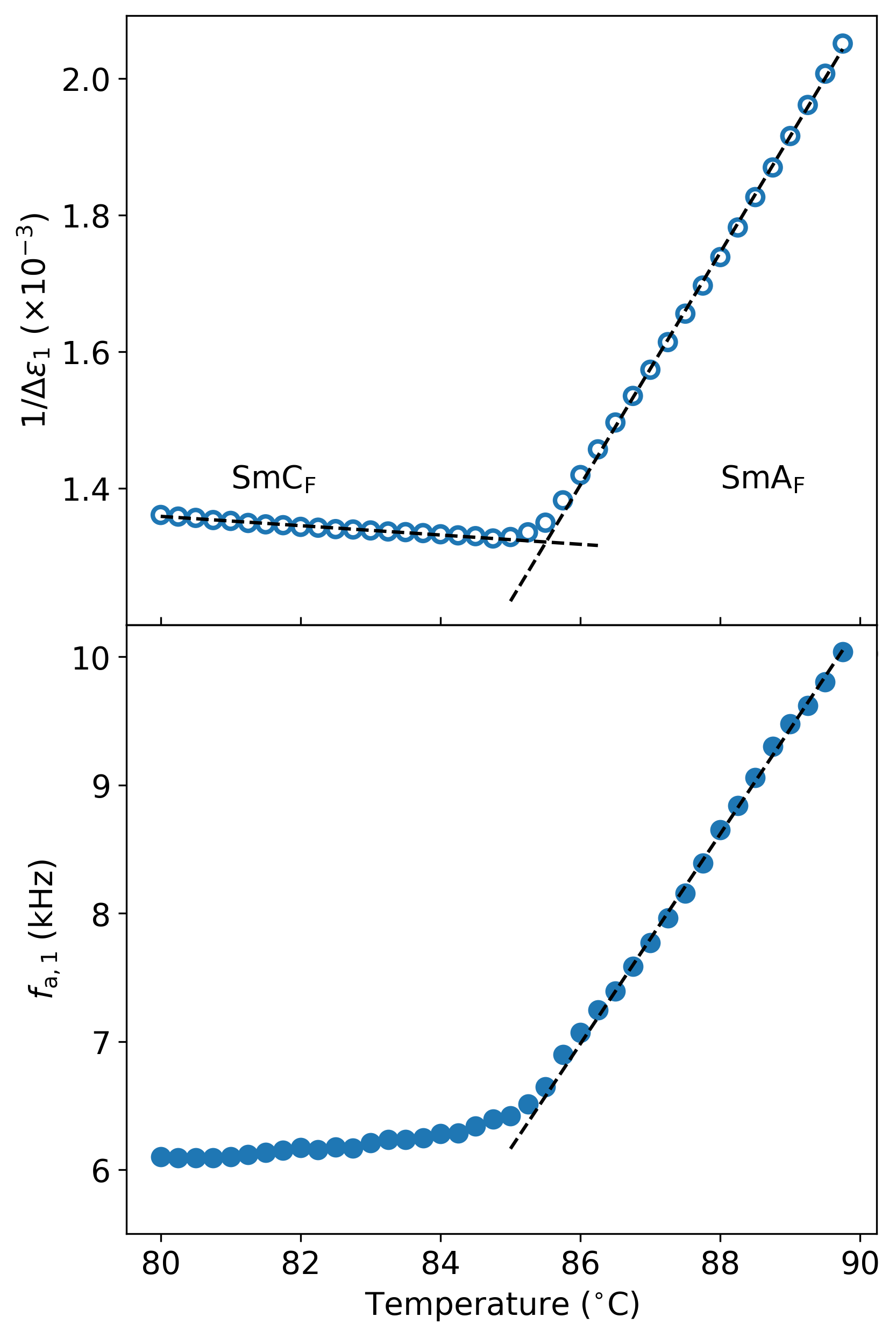}
\caption{\label{fig:figure6} Temperature dependence of the inverse of the dielectric strength ($1/\Delta \varepsilon_1$, top) and frequency of maximum absorption ($f_{\text{a,1}}$, bottom) of process m$_1$ at the SmA$_{\text{F}}$--SmC$_{\text{F}}$ transition. Black dashed lines are fits to the experimental data.}
\end{figure}

In the case of the SmA$_{\text{F}}$--SmC$_{\text{F}}$ transition, in contrast with the classical SmA--SmC transition, tilt angle fluctuations are coupled to the polar order (fluctuations in $|\psi|$ imply fluctuations in $|\mathbf{P}_{\perp}|$) so, along with the softening of the tilt elastic constant, we should observe the simultaneous divergence of the amplitude of the associated dielectric (soft) mode. If we analyze the data presented in Fig. \ref{fig:figure2} in detail, we see that it follows Curie-Weiss mean-field behavior $1/\Delta \varepsilon \propto (T-T_{\text{AC}})$, as shown in Fig. \ref{fig:figure6}. $1/\Delta \varepsilon$ obviously does not follow the classical prediction from the thermodynamic theory of ferroelectrics, according to which the slope of the low-temperature side should be twice that of the high-temperature one \cite{lines_principles_2001}. However, this also was not the case in classical ferroelectric SmC$^*$ LCs \cite{delafuente_low_1995, merino_molecular_1997, kremer_broadband_2003}. $f_{\text{a}}$ also nicely follows a linear dependence in the SmA$_{\text{F}}$ phase. The behavior in the SmC$_{\text{F}}$ phase may be explained by an increase in viscosity. For the classical SmA--SmC phase transition, of course, very little change was observed in dielectric measurements \cite{kresse_dielectric_1980}.

Regarding the microscopic mechanism of the phase transition, molecular dynamics simulations suggest that the origin of the
tilt is most likely the tendecy for staggered pairing which becomes
collective \cite{gibb_design_2026}. It would not be surprising that specific electrostatic interactions are behind this phenomenon, given that MIO is structurally almost identical to DIO and the latter does not exhibit either smectic phase. Moreover, our measurements reveal an unusually high bend elastic constant in the N phase, which indicates that the molecule is considerably straight and rigid or that the molecular interactions suppress bend deformations (or both). This then explains why MIO exhibits the SmC$_{\text{F}}$ phase and not the heliconical SmC$_{\text{P}}^{\text{H}}$ or N$_{\text{TBF}}$ phases \cite{gibb_spontaneous_2024, karcz_spontaneous_2024, erkoreka_flexoelectricity_2026}.

\section{Conclusions}

In this paper we have investigated the origin of molecular tilt in the ferroelectric smectic liquid crystal MIO. Despite its structural similarity with the prototypical ferroelectric nematogen DIO, this compound exhibits the phase sequence N--N$_{\text{S}}$--SmA$_{\text{F}}$--SmC$_{\text{F}}$, demonstrating how subtle molecular modifications can dramatically alter the balance between competing polar liquid-crystalline states.

Precision calorimetry has demonstrated that the transition is continuous and exhibits mean-field critical behavior. Light-scattering experiments have shown that the transition is driven by the softening of the tilt elastic constant. At the same time, dielectric spectroscopy reveals a Curie–Weiss-like divergence of the dielectric strength of the associated polar relaxation process, evidencing the intimate coupling between tilt and ferroelectric order. All together, these observations identify the SmA$_{\text{F}}$--SmC$_{\text{F}}$ transition as the ferroelectric counterpart of the classical SmA–SmC transition, modified by the presence of spontaneous polarization and long-range dipolar interactions. We especulate that the molecular rigidity of MIO and/or the presence of strong intermolecular correlations in the system, which strongly disfavor bend distortions, explain why the compound exhibits the SmC$_{\text{F}}$ phase as opposed to modulated SmC$_{\text{P}}^{\text{H}}$ or N$_{\text{TBF}}$ phases. In any case, further theoretical work through the development of microscopic models, or computational work through molecular dynamics simulations, will be required to establish this.

\section*{Acknowledgements}
A.E., I.A. and J.M.-P. acknowledge funding from the Basque Government Project IT1979-26 and from project PID2023-150255NB-I00 from MCIU/AEI/10.13039/5011000-11033/FEDER, UE. M.V.-A. acknowledges MCIU for his PhD grant (FPU24/01406). A.C. acknowledges grants PID2023-146811NA-I00 and RYC2021-031154-I funded by MICIU/AEI/10.13039/501100011033 and by the European Union NextGenerationEU/PRTR. A.G.-C. acknowledges support from María de Maeztu CEX 2023-001300-M/AEI/10.13039/501100011033.

\bibliography{REFERENCES}




\end{document}